\documentstyle[epsf]{l-aa}
\newcommand{\btab}{ \begin{tabbing}}
\newcommand{\etab}{ \end{tabbing}}
\newcommand{\diff}{{\rm d}}
\newcommand{\alphasynch}{\alpha_{\rm s}}

\newcommand{\kpar}{\kappa_{\scriptscriptstyle \|}}

\newcommand{\kperp}{\kappa_{\scriptscriptstyle \bot}}

\newcommand{\bdiff}{D_{\rm M}}

\newcommand{\eqb}{\begin{eqnarray}}
\newcommand{\eqe}{\end{eqnarray}}

\begin{document}
\thesaurus{02.04.2; 02.13.1; 02.16.1; 02.20.1; 11.03.4 Coma; 13.18.1}
\title{Anomalous transport of cosmic ray electrons}
\author{B. R. Ragot\inst{1,2} \and J. G. Kirk\inst{1}}
\institute{Max-Planck-Institut f\"ur Kernphysik,
Postfach 10 39 80, D--69029 Heidelberg,
Germany\and
Max-Planck-Institut f\"ur 
Radioastronomie, Auf dem H\"ugel 69, D--53121 Bonn, Germany
}
\offprints{J.G. Kirk}
\date{Received \dots Accepted \dots}
\maketitle

\begin{abstract}
Anomalous transport processes in which the variance of the distance 
travelled does not necessarily increase linearly with time: $\left<\Delta 
x^2\right>\propto t^{\alpha}$ with $0<\alpha<2$ are modelled using the 
formalism of continuous time random walks. We compute particle 
propagators which have the required dependence on space and 
time and use these to find the spatial dependence of the synchrotron 
radiation emitted by a population of continuously injected electrons. 
As the electrons are transported away from the source they 
cool, and the synchrotron spectrum softens. 
Sub-diffusive ($\alpha<1$) transport -- corresponding to stochastic 
trapping, or restriction of the transport across the average direction 
of a stochastic magnetic field -- produces a much 
slower rate of change of spectral index than does supra-diffusion 
($1<\alpha<2$) --
which occurs when particles move almost without scattering, in a field 
containing large ordered regions. Application to the diffuse emission 
of the outer parts of the Coma cluster favours an interpretation 
in terms of supra-diffusion. 
 
\end{abstract}

\section {Introduction}
\label{intro}
Many of the properties of cosmic rays can be understood on the basis of 
a propagation model in which they execute isotropic diffusion 
between injection at sources distributed in the galactic disk and 
escape at certain boundaries whilst possibly 
being carried out of the galaxy in a wind (Lerche \&
Schlickeiser~\cite{lercheschlickeiser82}
Berezinskii et al.~\cite{berezinskiietal90}, Webber et al.~\cite{webberetal92},
Bloemen et al.~\cite{bloemenetal93}). 
On the other hand, it has been known for 
many years that these particles are \lq magnetised\rq\ in the sense 
that their Larmor radius is much smaller than the mean free path
between \lq collisions\rq\ implied by the diffusion coefficient. This 
means that particles diffuse primarily along the magnetic field, and 
only very slowly across it. Since the galactic magnetic field lies 
predominantly in the galactic plane (e.g., 
Zweibel \& Heiles~\cite{zweibelheiles97}), it is surprising 
that isotropic diffusion describes cosmic ray propagation so well.
One might instead expect cosmic rays to be transported anomalously 
slowly in the direction normal to the disk 
(Getmantsev~\cite{getmantsev63}). The success of the 
simple diffusion model of propagation (or even of the more primitive 
but related \lq leaky-box\rq\ model) appears to show that this is not the 
case, at least for the bulk of cosmic rays. This is presumably due to 
fluctuations in the magnetic field, possibly connected with 
instabilities, which cause a given field line to wander out of the 
galactic plane, and give rise to an effective diffusion across the
direction of the mean magnetic field (Jokipii \& 
Parker~\cite{jokipiiparker69}, Jokipii~\cite{jokipii73}).

Nevertheless, it may not always be the case that the transport 
process can be described by an effective diffusion coefficient. 
Since its first mention by Getmantsev~(\cite{getmantsev63}), 
anomalous transport, i.e., transport in which the mean square deviation 
of a particle from its position at a given time 
$\left<\Delta {\bf x}^2\right>$ does not depend linearly on the 
elapsed time, has been discussed in two astrophysical contexts. 
Firstly, Chuvilgin \& Ptuskin~(\cite{chuvilginptuskin93}) derived a 
kinetic equation for cosmic ray transport for the case in which 
$\left<\Delta {\bf x}^2\right>\propto t^{1/2}$, but pointed out that 
this type of transport can be expected only when considering timescales 
short compared to the time taken for a particle to effectively 
decorrelate from a given magnetic field line. 
Many effects such as particle drifts, temporal
changes in the magnetic field or just the chaotic structure of the 
field itself
can cause decorrelation, but although it is very difficult 
to estimate this time in an astrophysical situation, the success of 
the diffusion model suggests that cosmic rays do indeed decorrelate from the 
field faster than they escape from the galaxy (which takes about 
$10^8\,$years). 

The second astrophysical context is that of particle acceleration at 
a shock front. In this case, the natural timescale is the 
acceleration time, which is thought to be a strong function of energy, 
and to vary over several orders of magnitude for those cosmic rays 
accelerated at supernova shocks (Dendy et al~\cite{dendyetal95},
Duffy et al.~\cite{duffyetal95}). 
Here too, it is difficult to make a 
realistic estimate of the decorrelation time of a particle from the 
magnetic field. However, predictions can be made of the spectrum and 
spatial distribution of those particles undergoing anomalous 
transport, which might enable them to be distinguished from diffusing 
particles (Kirk et al.~\cite{kirketal96a}).

A problem which is similar in many respects arises in the confinement of plasma 
in fusion devices, and in this context there has been considerable 
interest in recent years in anomalous, non-diffusive transport models
(Rechester \& Rosenbluth~\cite{rechesterrosenbluth78},
Kadomtsev \& Pogutse~\cite{kadomtsevpogutse79},
Isichenko~\cite{isichenko91a},\cite{isichenko91b},
Rax \& White~\cite{raxwhite92}, Wang et al.~\cite{wangetal95}). 
In particular, use 
of the formalism developed for continuous time random walks (CTRW -- 
see Montroll \& Weiss~\cite{montrollweiss65})
has been 
advanced (Balescu~\cite{balescu95}). This approach is 
widely used in the problem of anomalous transport in random media 
(e.g., Bouchaud \& Georges~\cite{bouchaudgeorges90}, 
Shlesinger et al.~\cite{shlesingeretal93}).
In this paper, we discuss these techniques 
and show how they relate to the work already done on anomalous transport in 
astrophysics. We point out that they offer a more general approach to the 
problem, enabling one to relax the rather restrictive assumptions
concerning the statistical properties of the magnetic field 
employed hitherto. As an 
example, we examine a new astrophysical application -- 
that of the synchrotron emission of a population of relativistic electrons 
which lose energy whilst being transported away from the site of their 
acceleration. Both the spatial extent of the radiating electrons and 
the observed spectrum depend on the nature of the transport process. 
These calculations are especially relevant to the interpretation 
of high resolution radio observations in several 
astrophysical situations, including 
spiral galaxies seen \lq edge-on\rq\ 
(e.g., Hummel et al.~\cite{hummeletal91}) as well 
the diffuse emission from clusters of 
galaxies (Schlickeiser et al.~\cite{schlickeiseretal87}, 
Kirk et al.~\cite{kirketal96b}).
\section{Anomalous transport}
In a regular magnetic field topology, 
the transport of charged particles across the field 
is due to the collisions of the particles and their 
finite Larmor radii. However, a perturbation of the 
magnetic field results in a wandering of the field 
lines and a potentially much faster transport of the particles. 
This transport is widely attributed to the effect 
of the large-scale random field component, 
which would produce, following the quasilinear 
theory, a diffusion of the field lines across the direction of the the average
field (Jokipii \& Parker~\cite{jokipiiparker69}, Kadomtsev \&
Pogutse~\cite{kadomtsevpogutse79}). In fusion applications, it is generally 
assumed that particle collisions lead to diffusive transport along each
individual field line. 
In astrophysical applications, where collisions can be neglected, it is usual
to assume that small scale fluctuations in the magnetic field play this
role, so that here too, particles diffuse along field lines (Chuvilgin \&
Ptuskin~\cite{chuvilginptuskin93}). 
As long as the particles remain correlated to a given 
patch of field lines, the combination of these two diffusions 
results in sub-diffusion of the particles i.e., $\langle 
\Delta x^2(t) \rangle \propto t^{\alpha}$, with $\alpha = 1/2$, 
as described by Getmantsev~(\cite{getmantsev63}). 
(In the general case, sub-diffusion refers to all such processes when 
$0 < \alpha < 1$.)
However, the exponential divergence of neighbouring field lines 
and the resulting stretching of a field 
line patch lead to decorrelation of a 
particle from its field line, and thus to the large-scale diffusion 
of the particles, as pointed out by 
Rechester and Rosenbluth~(\cite{rechesterrosenbluth78}), 
with a diffusion coefficient given by the  
expression: 
\eqb
D_{RR} &=& D_{st} 2 \kpar/l_{c\delta}
\eqe
where $l_{c\delta} = l_c \log(1/k_0 \delta)$,   
$\delta = l_c \sqrt{\kperp/\kpar}$, where  
$l_c$
is the exponentiation length of the field lines. 
Here 
$D_{st}$ is the quasilinear diffusion coefficient of the field lines, 
which has the dimensions of a length, 
$k_0$ is a characteristic wave number of the magnetic turbulence, $\kpar$ 
and $\kperp$ are the quasilinear diffusion coefficients of the particles 
along and across the direction of the local magnetic field, 
respectively.
The transport of particles across a turbulent magnetic field is 
thus sub-diffusive for short timescales, but crosses 
over to normal diffusion in the limit of the long timescales. 
In the following, we investigate the anomalous transport regime, which 
applies if the natural timescale 
of a particular problem -- such as 
escape from the galaxy, acceleration at a shock, or loss of energy by
synchrotron radiation, is shorter than the decorrelation time.

In a statistically homogeneous medium, 
the density of particles, $n(\vec{x},t)$, is related to the 
source $Q(\vec{x}',t')$ by the propagator $P$:
\eqb
n(\vec{x},t) &=& \int \diff\vec{x}' \int \diff t' \, P(\vec{x}-\vec{x}',t-t') 
Q(\vec{x}',t') \; .
\label{density}\eqe
For diffusive transport, $P$ typically contains a
factor $\exp(-|\vec{x}|^2/Dt)$, and is the Green's function of the
diffusion equation.

For sub-diffusive transport with $\alpha=1/2$, Rax \& 
White (\cite{raxwhite92}) have determined the propagator 
by combining two diffusive propagators using a Wiener integration
method. 
They considered the transport in cylindrical symmetry across the
$z$ direction (i.e., a 2-dimensional problem, with
$|z|$ playing the role of time), and obtained: 
\eqb
P(r,t) &=& \frac{H(t)}{2\pi\left( 4 r^{2} \bdiff^{2}
\kpar t\right)^{1/3}} \exp\left[ \frac{-3 r^{4/3}}
{4 \left(4  \bdiff^{2} \kpar t\right)^{1/3}} \right]\,    
\label{prop2}
\eqe
where $r$ is the radius in cylindrical coordinates, and $H(t)$ the 
Heaviside function.
Duffy et al.~(\cite{duffyetal95}) considered the transport 
perpendicular to a plane shock front. In a cartesian coordinate system
with the shock in the $y$--$z$ plane, and assuming no gradients in the 
$y$ direction are present, they obtained the propagator 
\eqb   
P(x,t) &=& \int \diff z \, \frac{\exp\left[-x^2 / \left(4 \bdiff |z|\right)
-z^2 / \left(4 \kpar t\right)\right]}
{\sqrt{4 \pi \bdiff |z|}\sqrt{4 \pi \kpar t}}
\label{doublediff}
\eqe
Approximating the integral using the method of steepest descents, they found
\eqb
P(x,t)&\approx& 
\eta (\bdiff \kpar^{1/2} |x| t^{1/2})^{-1/3} 
\exp \left( \frac{-\beta |x|^{4/3}}{\bdiff^{2/3}\kpar^{1/3} 
t^{1/3}} \right) 
\label{prop1} 
\eqe
with $\eta = 2^{1/3} (3\pi)^{1/2}$ and $\beta = 3/2^{8/3}$.
For this 1-dimensional problem of $\alpha=1/2$ sub-diffusion,
Chuvilgin \& Ptuskin~(\cite{chuvilginptuskin93}) derived an
equation describing the evolution of the particle density 
(their Eq.~B.12) and solved it 
to find the above propagator. Essentially the same equation was also
found by Balescu~(\cite{balescu95}). It can be written: 
\eqb
\partial_t n(x,t) &=&  D_0 \Bigg[\nabla_x^2 n(x,t)
\nonumber\\
&&
-{1\over2\sqrt{\pi}\tau_D}
\int_{1/\pi}^t d\tau \left(\frac{\tau_D}{\tau}\right)^{3/2} 
\nabla_x^2 n(x,t-\tau) \Bigg] \; , 
\label{nmde}
\eqe
with $ D_0 = \sigma^2/2 \tau_D$.
This is a non-Markovian diffusion equation: the integral term is 
characteristic of the long-time memory of the dynamics. 

In a realistic situation, the topology of the magnetic 
field may be more complicated 
than just a pure stochastic sea with its associated
diffusion of the field lines. In fusion plasmas, for example, 
there can exist ordered structures, \lq stability islands\rq, 
in which particles can be trapped for long periods of time, 
(referred to as \lq sticking\rq), leading to sub-diffusive large-scale 
transport of the field lines themselves (e.g., 
White et al.~\cite{whiteetal93}).
On the other hand, the field lines may also wander 
faster than implied by diffusion. In this case, the field lines are said to
perform \lq flights\rq, during which they maintain almost the same direction 
for a relatively long distance. 
Ultimately, the large-scale transport of field lines 
is the result of competition between \lq sticking\rq\ and \lq flights\rq,  
and can yield transport regimes of many different kinds, 
ranging from slow sub-diffusion (almost perfect sticking) 
to fast supra-diffusion (domination of flights). In terms of 
 $\alpha$, the physically relevant range is $0<\alpha <2$, with
$\alpha=2$ corresponding to transport completely dominated
by a single straight-line flight. 

To find the particle propagator, it is necessary to combine 
the propagator for field lines with that for particle motion along
the field, as described above for the case of $\alpha={1\over2}$.
Here too, the assumption of diffusive transport can be generalised. Particles 
may, in fact, undergo no scattering at all, in which case they propagate 
ballistically along the field lines, as assumed by 
Achterberg \& Ball~(\cite{achterbergball94}). On the other hand, they may be
trapped between magnetic mirrors on a segment of a field line. Once again, the
transport can be characterised by an index $\alpha$ which lies between 0 and
2. However, as we show below, the combination of two propagators does not
extend the overall range  of $\alpha$ which is permitted. In fact, 
the effect of superimposing transport with the two indices $\alpha_1$ and 
 $\alpha_2$ is described by a single value $\alpha=\alpha_1\alpha_2/2$.

\section{Continuous time random walks}
\label{ctrwsect}
We do not aim to describe the microscopic dynamics of particles 
moving in a turbulent magnetic field in a realistic way, but rather look
for a  model which is able to reproduce their transport 
characteristics and gives the density profile.
In terms of the propagator $P(\vec{x},t)$, 
the density is given by
Eq.~(\ref{density}), once the rate $Q(\vec{x},t)$ at which 
particles are injected is specified.
This can then be used to compute, for example, the synchrotron radiation
from electrons injected at a plane surface (such as a spiral galaxy) and 
subsequently transported outwards. 

The method we adopt to model $P$ is that of continuous 
time random walks (CTRW). In this, the transport of a
particle is modelled as a succession 
of instantaneous jumps of arbitrary length, 
separated by pauses of arbitrary duration. Each of these steps is 
independent of the previous steps. We denote by $f(\vec{x})$ 
the probability distribution function (PDF) of a jump described by 
$\vec{x}$, and by $\psi(t)$ the PDF of a pause of duration $t$, 
and use their Fourier and Laplace transforms: 
\begin{equation}
\tilde{f}(\vec{k}) = \int d\vec{x} \, f(\vec{x}) e^{i\vec{k}.\vec{x}} 
\; \; ; \; \; \; \hat{\psi}(s) = \int_0^{+\infty} dt \, \psi(t) e^{-st} \; . 
\label{FLT}
\end{equation} 

The type of anomalous transport characterised by $\alpha$ and discussed above
arises from the asymptotic behaviour of $P(\vec{x},t)$ as $|\vec{x}|  
\rightarrow + \infty$ and $t \rightarrow +\infty$. This corresponds  
to the $|\vec{k}| \rightarrow 0$, $s\rightarrow 0$ asymptotic behaviour of
 $\hat{\tilde{P}}(\vec{k},s)$, for which the expansions of $\hat{\psi}(s)$ 
and $\tilde{f}(\vec{k})$ around 0 are required. 
To keep the treatment general, we consider first 
the motion of a particle in a space 
of dimension $d$. 
The functions  $\hat{\psi}(s)$ 
and $\tilde{f}(\vec{k})$ are then prescribed as
\begin{eqnarray}
\hat{\psi}(s) &=& 1-\tau_D^{\alpha} s^{\alpha} \; , \; \;  
s \rightarrow 0 \\ 
\tilde{f}(\vec{k}) &=& 1 - \frac{1}{2d} \sigma^{\beta} |\vec{k}|^{\beta} 
\; , \; \; |\vec{k}| \rightarrow 0 \; ,  
\label{EXP} 
\end{eqnarray} 
with $0< \alpha, \beta \leq 2$, and $\tau_D$, $\sigma$ 
positive 
constants. In Eq.~(\ref{EXP}), we have assumed that $\tilde{f}$ depends only on
$|\vec{k}|$,
i.e., that there is no preferred direction for a particle jump. 
The particular values 
$\alpha=1$, $\beta = 2$ yield a diffusive process with
a Gaussian propagator, whereas the case $\beta < 2$, leads to a random walk
with 
an infinite second moment $\langle x^2 \rangle$, corresponding to L\'evy
flights.
   
In a classical paper, Montroll and Weiss~(\cite{montrollweiss65}) have shown 
that the
Fourier-Laplace transform of $P(\vec{x},t)$ has the following form
\begin{equation}
\hat{\tilde{P}}(\vec{k},s) = \frac{1-\hat{\psi}(s)}
{s[1-\hat{\psi}(s) \tilde{f}(\vec{k})]} \; .
\label{MWE}
\end{equation}
Inserting the expansions Eq.~(\ref{EXP}) we find that for large $|\vec{x}|$ and
$t$
\eqb
\langle |\vec{x}|^2 \rangle	& =& \int_{-\infty}^{+\infty} d\vec{x} 
\, |\vec{x}|^2 P(\vec{x},t)\nonumber\\
& =& 
\frac{1}{2\pi i} \frac{1}{2\pi} \int_{\Gamma} \diff s \int_{-\infty}^{+\infty} 
\diff \vec{k} \int_{-\infty}^{+\infty}\diff\vec{x} \, 
\nonumber\\
&&
\exp\left(st-i\vec{k}.\vec{x}\right)
|\vec{x}|^2 
\frac{\tau_D^{\alpha} 
s^{\alpha-1}}{\tau_D^{\alpha}s^{\alpha}+\sigma^{\beta}|\vec{k}|^{\beta}/(2d)} 
\nonumber \\ 
&&\propto  t^{2\alpha / \beta} 
\label{GSM} 
\eqe
and the transport coefficient is $\mu = 2\alpha / \beta$. 
In the following, we find it convenient for the 
inversion of the Fourier transforms to choose $\beta=2$, 
and describe all transport 
regimes using the parameter range $0<\alpha\le2$:
\eqb
\hat{\psi}(s) & = & 1-\tau_D^{\alpha} s^{\alpha} \; , \; \; 
0 < \alpha \leq 2 \\  
\tilde{f}(\vec{k}) & = & 1 - \frac{1}{2d} \sigma^2 |\vec{k}|^2 \; , 
\label{EXPS} 
\eqe
for small $s$ and $|\vec{k}|$.
The choice $\alpha = 1$ has the same asymptotic behaviour as the function 
$\hat{\psi}=\exp(-s/\tau_D)$, i.e., $\psi=\delta(t-\tau_D)$, which 
corresponds to scattering after a fixed time interval, yielding 
diffusion. However, for 
$\alpha\ne1$,$2$, we have
\eqb
  \psi (t) &=& \frac{1}{\tau_D} \frac{\alpha}
{\Gamma(1-\alpha)} \left(\frac{t}{\tau_D}\right)^{-1-\alpha}
\eqe 
for $t \rightarrow +\infty$. This is a long-tailed distribution 
function. Its slow decrease at large $t$ leads to the same kind of 
memory effect built in to 
the non-Markovian transport equation (\ref{nmde}) 
constructed by Chuvilgin \& Ptuskin~(\cite{chuvilginptuskin93}) 
and Balescu~(\cite{balescu95}) for the case $\alpha<1$.  
We do not consider $\alpha>2$, since in this case the quantity
$\left<\vec{x}^2\right>/t^2$ increases without limit for large $t$, 
and would at some stage exceed the actual physical particle speed.

From the expansions (\ref{EXPS}) the asymptotic behaviour of the 
propagator at large $|\vec{x}|$ and $t$ can be found by an inverse Fourier
transformation followed by an inverse Laplace transformation, which is
performed approximately using the method of steepest descents. The resulting
expression can be used for all values of $\vec{x}$ and $t$ once the 
normalisation has
been corrected. It represents a convenient one-parameter model of particle
transport which exhibits the anomalous behaviour in which we are interested.
The details of the derivation are given in Appendix~A. 
Here we simply present the results,
expressed in terms of a dimensionless similarity variable defined according to 
\eqb
\xi&=&\sqrt{2 d}{|\vec{x}|\over\sigma} \left({\tau_D\over t}\right)^{\alpha/2}
\label{xidef}
\eqe
Using this variable, the propagator separates:
\eqb
P_d(\vec{x},t)&=& \sigma^{-d}\left({\tau_D\over t}\right)^{\alpha d/2} \Xi_d(\xi)
\enspace,
\label{propsep}
\eqe
and we find
\eqb
\Xi_d(\xi)&=& C_d \xi^{-d(1-\alpha)/(2-\alpha)}
\exp\left[-A \xi^{2/(2-\alpha)}\right]
\label{xioned}
\enspace,
\eqe
where the (positive) constants $C_d$ and $A$ are given in the appendix
(Eqs.~\ref{constants1} and \ref{constants2}).
These propagators are normalised such that particle number is conserved: 
$\int {\rm d}^d\vec{x}P(\vec{x},t)=1$. They display at all $t$ the 
behaviour required of the asymptotic dependence of the 
variance in anomalous transport. In particular, they possess for all $t$ 
the property
\eqb
\left<x^\nu\right>\propto t^{\nu\alpha/2}
\enspace.
\eqe
  
For one-dimensional propagation ($d=1$), which describes, for example, 
the transport
of particles in $x$ away from a uniform source in the $y$--$z$ plane, we
have for the standard diffusive case ($\alpha=1$) the well-known form
\eqb
P(x,t) &=& \frac{1}{\sqrt{4\pi Dt}} \exp\left[-\xi^2/4\right] 
\label{DPROP} 
\eqe 
with $D=\sigma^2/(2\tau_D)$ and $\xi=|x|/\sqrt{D\,t}$.
In the $\alpha=1/2$ sub-diffusive case, which can be considered as
a double diffusion, we find: 
\eqb
P(x,t) &=& \frac{1}{\sigma\sqrt{6\pi}}\left({\tau_D\over
t}\right)^{1/4}\xi^{-1/3}
\exp\left[-3\left(\xi/4\right)^{4/3} \right]
\label{SDPROP} 
\enspace,
\eqe
which agrees with the propagator for $d=1$ given in   
Eq.~(\ref{prop1}) and derived by Duffy et al.~(\cite{duffyetal95}), with
$\sigma = (2D_M)^{2/3} k_\|^{1/3} / u^{1/3}$ 
and $\tau_D=(2D_M)^{2/3} k_\|^{1/3} / u^{4/3}$.
(Similarly, the case $d=2$ reproduces the propagator (\ref{prop2}) of 
Rax \& White~\cite{raxwhite92}). 
In analogy with the formulation of 
Eq.~(\ref{doublediff}), it is possible to consider other values of $\alpha$ as
arising from a convolution of a propagator describing motion along the field
line together with one describing the wandering of the field. Using the 
method of steepest descents, it is straightforward to show that $P_{d,\alpha}$ 
has the following convolution property: 
\begin{equation} 
\int_{-\infty}^{+\infty} ds \, P_{d,\mu_1}(\vec{x},|s|) \, 
P_{1,\mu_2}(s,t)  = P_{d,\mu_1 \mu_2/2}(\vec{x},t) 
\label{combi} 
\enspace,
\end{equation} 
(note that the normalisation is preserved)
so that the combined transport process can always be described using a value of
$\alpha$ between 0 and 2.

\begin{figure}
\epsfxsize=8 cm
\epsffile{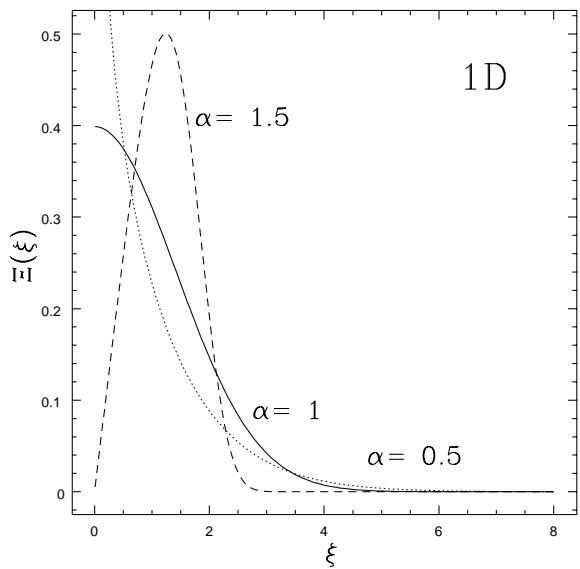}
\caption{\label{propfig}
The dimensionless \hbox{propagator} 
(Eq.~\protect\ref{propsep}) 
for one-dimensional transport in three different regimes:
standard diffusion ($\alpha=1$, solid line), sub-diffusion ($\alpha=1/2$ dotted
line) and supra-diffusion ($\alpha=3/2$ dashed line), as a function of the
dimensionless similarity variable $\xi$, which, for fixed time, 
is proportional to distance (see
the definition in Eq.~\protect\ref{xidef}).}
\end{figure}

To illustrate the properties of these propagators, they are plotted
for three different values of $\alpha$ in Fig.~(\ref{propfig}).
The diffusive propagator is simply a gaussian curve given by Eq.~(\ref{DPROP})
according to which the probability of finding a particle at a particular
distance drops off smoothly with increasing distance from the point (plane) of 
injection. The sub-diffusive propagator, on the other hand, falls off much more
steeply initially (it has an integrable singularity at the point of injection),
but at larger distances decreases more slowly than diffusion. 
Acting on an
ensemble of particles, sub-diffusive transport tends to confine some of them
close to the injection point, but propagates others very rapidly to large
distance. This is a result of the larger spread in position of sub-diffusing 
particles of a given age, compared with diffusing ones. Thus, despite the fact
that sub-diffusion produces, on average, slower transport, a minority of 
particles experiences comparatively rapid transport.
Supra-diffusion displays the opposite behaviour and is similar to ballistic
transport, or propagation at constant speed without scattering. The propagator
is strongly peaked around the value $\xi\approx1$. Very few particles are to
be found lingering close to the point of injection, and very few escape to
large distance.

Equation~(\ref{xioned}) is the basic result of this section.
We now turn to the question of the observable consequences of these
propagators.

\section{Synchrotron radiation from anomalously transported electrons}
The synchrotron radiation emitted by relativistic electrons is the most important diagnostic available for the study of the transport of these particles in astrophysical environments. In this section we compute the spatial dependence of the surface brightness and of the spectral index which is to be expected when the different kinds of particle transport discussed above dominate.

The energy losses suffered by a relativistic particle emitting synchrotron radiation, or undergoing inverse compton scattering whilst propagating through a homogeneous magnetic field, or a homogeneous soft radiation field can be described by the equation
\eqb
{\diff \gamma\over\diff t}&=&-g\gamma^2
\label{lossrate}
\eqe
where $\gamma$ is the Lorentz factor of the particle and 
$g$ a factor which depends on the magnetic 
field strength, and the energy density of the photon field.
In units suited to the application to clusters of galaxies (see below) one has
\eqb
g= 4.1\times10^{-14}
\left( B^2_{\mu \rm G} + 12.3\right)\qquad {\rm yr^{-1}}
\eqe
where the photon field is assumed to be that of the cosmic microwave 
background, 
which is for this purpose equivalent to a magnetic field strength of $3.5\,{\rm \mu G}$. Defining the differential rate $\diff Q$ at which electrons are injected into the system at time $t$ as $\diff Q=Q(\vec{x},\gamma,t)\diff^d\vec{x}\diff\gamma$, the resulting electron density, allowing for energy losses, is
\eqb
n(\vec{x},\gamma,t)&=&
\int_\gamma^\infty \diff\gamma'\int\diff^d\vec{x'}\int_{-\infty}^t\diff t'
P(\vec{x}-\vec{x'},t-t')\nonumber\\
&&Q(\vec{x'},\gamma',t')
\delta\left(\gamma-{\gamma'\over 1+g\gamma'(t-t')}\right)
\enspace.\label{genexprden}
\eqe
Consider electrons continuously injected with a power-law distribution in 
$\gamma$ at the point $\vec{x}=0$, i.e.,
\eqb
Q(\vec{x},\gamma,t)&=&Q_0\gamma^{-p}\delta(\vec{x})
\label{injspectrum}
\enspace,
\eqe
where $Q_0$ is a constant. At all points and at all energies of interest the distribution has had sufficient time to reach a stationary state, which is given by the integral
\eqb
n(\vec{x},\gamma)&=&Q_0\gamma^{-p}\int_0^{1/(g\gamma)}
\diff t' P(\vec{x},t')(1-g\gamma t')^{p-2}
\label{statdens}
\enspace.
\eqe
In order to compute the surface brightness on the sky of the 
synchrotron radiation from these electrons, two more integrations are 
required -- firstly over the line of sight through the source and secondly 
over $\gamma$, after multiplying $n$ by the synchrotron kernel. Furthermore, 
information about the magnitude and 
direction of the magnetic field along the line of sight 
is needed. 
For our purpose, it is sufficient to use the delta-function or 
\lq monochromatic\rq\ approximation to the synchrotron kernel 
(see, for example, 
Mastichiadis \& Kirk~\cite{mastichiadiskirk95})
and we will simplify the discussion by using a constant, angle-averaged
emissivity in the line of sight integral 
(see, however, Crusius \& Schlickeiser~\cite{crusiusschlickeiser88}). 
As a result, the surface brightness at a given frequency is 
simply proportional to $\int\diff z n(\vec{x},\gamma)$, where $z$ is 
measured along the line of sight and where $\gamma$ is fixed, being 
proportional to the square root of the observing frequency. 

\begin{figure}
\epsfxsize=8 cm
\epsffile{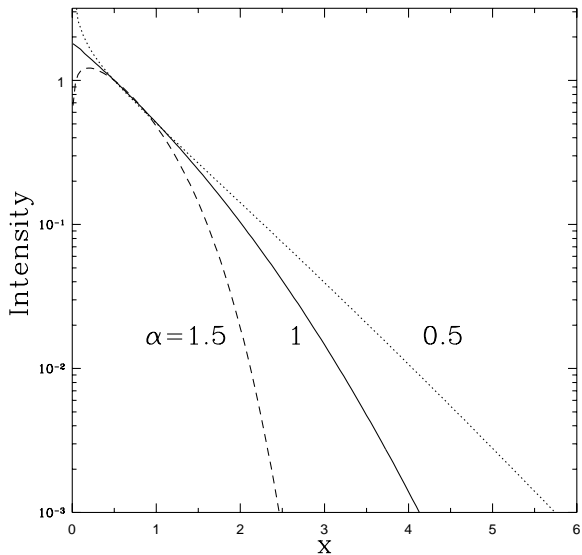}
\caption{\label{1dprofile}
The surface brightness of synchrotron radiation from a planar source of
relativistic electrons, seen edge-on, as a function of distance $x$ from
the plane of injection, normalised to unity at $x=0.5$. Distance is 
expressed in units of the scale length $x_s$ (Eq.~\protect\ref{xscale}).  
Electrons are injected at $x=0$ with a spectrum 
$Q\protect\propto\protect\gamma^{-2.5}$
}
\end{figure}

In the one-dimensional case in which electrons are injected in a plane, 
and the line of sight is parallel to this plane, displaced by a distance 
$x$, the resulting surface brightness is directly proportional to 
 $n(x,\gamma)$ given by Eq.~(\ref{statdens}). This expression is simple to 
evaluate numerically. For various special cases, such as diffusion
($\alpha=1$), or injection with $p=2$ the integral can be written in 
closed form (see Appendix~B for details). The resulting profile is shown in 
Fig.~\ref{1dprofile} for the three qualitatively different types of transport considered in Sect.~\ref{ctrwsect}.
It is important to note that the shapes of 
these profiles are characteristic of
the type of 
transport involved and are independent of the frequency of observation. 
The assumptions entering the calculation are that the magnetic 
field strength is constant, that the source has the stated geometry, and that
the electron injection has been steady over a sufficiently long period of
time. The abscissa in Fig.~\ref{1dprofile} is the scaled radius $\hat{x}=x/x_s$, where
\eqb
x_s&=&{\sigma\over\sqrt{2}(g\gamma\tau_D)^{\alpha/2}}
\label{xscale}
\enspace,
\eqe
which is in general a function of observing frequency. 
Sub-diffusive transport ($\alpha<1$),
which tends to trap particles close to the point of injection, displays an
integrable singularity at $x=0$, whereas supra-diffusive transport, which
quickly moves particles away from the source, initially increases at small
$x$. The brightness profile for diffusive transport is monotonically decreasing
with $x$. In Fig.~\ref{1dprofile} the surface brightness is arbitrarily
normalised to unity at $x=0.5$.
In a realistic situation, in which the electron source is distributed in space,
it may not be possible to detect these features. In fact, the smooth
almost exponential decrease of surface brightness with distance means that it
would be difficult if not impossible to distinguish observationally between the different types
of transport from these profiles alone.  

\begin{figure}
\epsfxsize=8 cm
\epsffile{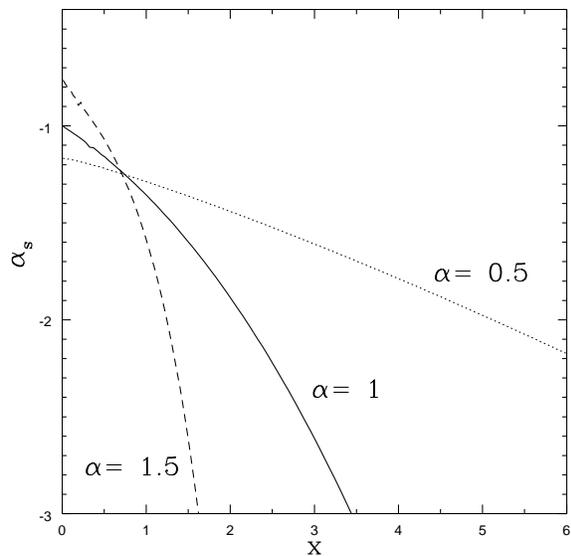}
\caption{\label{1dindexplot}
The spectral index of synchrotron emission as a function of (scaled) 
radius. The dotted line corresponds to sub-diffusion with 
$\protect\alpha=0.5$, the
solid line to diffusion ($\protect\alpha=1$) and the dashed line to 
supra-diffusion
($\protect\alpha=1.5$). Electrons are injected continuously in 
the plane $x=0$ with a
spectrum $Q\protect\propto\protect\gamma^{-2.5}$. 
The total emission integrated over
 $x$ is thus a pure power-law of index $\protect\alphasynch =-1.25$. 
}
\end{figure}

A more promising approach lies in the measurement of synchrotron
spectra as a function of position. In the
delta-function approximation, the synchrotron spectral index $\alphasynch$ is
simply related to the $\gamma$-dependence of the 
electron density integrated along the line of sight. 
To compute this, one must take
account of the dependence of the scaling radius on $\gamma$. 
Defining the frequency dependence of the surface brightness 
by $I_\nu\propto\nu^{\alphasynch}$, we have
\eqb
\alphasynch&=&- {1\over2}\left\lbrace p-{(\alpha)\over2}\left[
1+
\left({\partial \log(N)\over\partial\log(\hat{x})}\right)\right]
\right\rbrace
\label{synchindex}
\eqe
where $N=\int\diff z\, n(\vec{x},\gamma)$. The resulting spectral index as a
function of $x$  is plotted in Fig.~\ref{1dindexplot} for $p=2.5$.

This figure shows that the different types of transport produce very different
spectra as the particles propagate away from the source. 
Close to the plane of injection, supra-diffusion
gives approximately the \lq\lq uncooled\rq\rq\ value of the spectral index
$\alphasynch=-0.75$. This is because only recently injected particles remain
close to $x=0$ -- older particles have a very small chance of return in the
supra-diffusive case. As $\alpha$ decreases, more and more older particles are
confined close to the point of injection, so that the spectrum for diffusive
and sub-diffusive transport is softer. In general, the age distribution of
particles produced by each transport process determines the spectrum. Thus,
supra-diffusion ($\alpha=1.5$, dashed line) has a very narrow spread of
particle age at a given $x$. Once the distance exceeds $x_s$, the spectrum
becomes very soft, since the number of radiating particles decreases rapidly.
In the case of diffusion ($\alpha=1$, solid line) the age distribution is
gaussian and produces a more gradual softening, whereas the
sub-diffusive case ($\alpha=0.5$, dotted line) not only traps some particles
close to $x=0$ but, also transports a few
particles out rapidly outwards. Thus, the spectrum remains rather hard 
 even at large radius. An observationally relevant measure of this dependence is
the change in spectral index over the distance required for the brightness to
drop by one order of magnitude from its value at $x=0.5$. For supra-diffusion,
the spectrum softens by $1.8$ from $-1$ at $x=-.5$ to $-2.9$ at $x=1.6$; for
diffusion it softens 
by $0.8$ from $-1.1$ at $x=0.5$ to $-1.9$ at $x=2$
and for sub-diffusion we find a change of only $0.3$ from $-1.2$ at $x=0.5$ to
$-1.5$ at $x=2.2$. Thus, a relatively slow change
in synchrotron spectral index indicates sub-diffusive behaviour.

As an example, we consider the diffuse radio emission observed from the 
Coma cluster of galaxies. There is active debate as to the origin
of the relativistic electrons responsible for this emission. It appears
that simple diffusion from one of the galaxies 
near the centre is
not, by itself, a viable explanation (e.g., Kirk~et al.~\cite{kirketal96b})
and that some kind of distributed 
injection and/or acceleration is necessary (Schlickeiser et al.~\cite{schlickeiseretal87}). However, observations by 
Giovannini et al.~(\cite{giovanninietal93}) 
have shown that there is a distinction between the central 
parts of the cluster 
(say within a radius of $500\,{\rm kpc}$) and the outer 
parts (in this paper we adopt the value $H_0=50\,{\rm km\,s^{-1}\,Mpc^{-1}}$, 
so that 1~sec of arc is equivalent to 700~pc). If we assume that the processes
of injection and acceleration occur 
within a sphere of radius $500\,{\rm kpc}$, we can apply the above propagators
to model the surface brightness and spectral index of the emission in the 
outer parts. 

\begin{figure}
\epsfxsize=8 cm
\epsffile{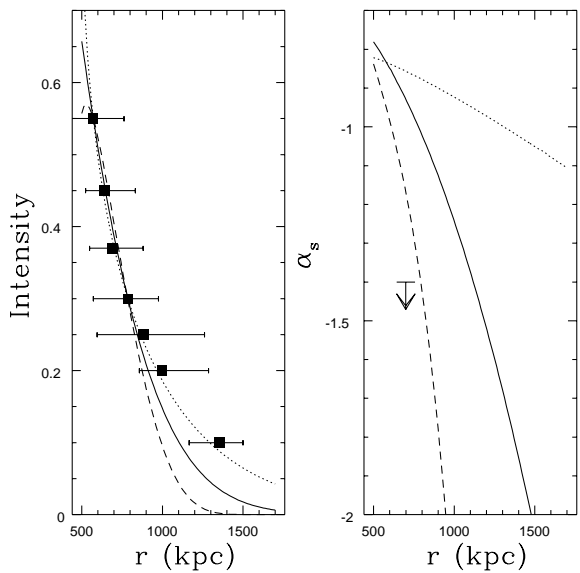}
\caption{\label{coma1d}The 
one-dimensional surface brightness profiles for the three types of transport
$\alpha=0.5$, $1$ and $1.5$ fitted to observations of the diffuse 
radio emission of the Coma cluster of galaxies by
Hanisch~(\protect\cite{hanisch80}) 
as presented by Valtoja~(\protect\cite{valtaoja84}). Only the outer parts
of the cluster are plotted. Also shown are the predicted dependences of 
the spectral index on radius. The upper limit is a rough indication of the 
spectral softening observed by Giovannini et al.~(\protect\cite{giovanninietal93})
}
\end{figure}
 
Figure~\ref{coma1d} shows the observed emission at $430\,{\rm MHz}$, as 
presented by Valtaoja~(\cite{valtaoja84}), from the data of 
Hanisch~(\cite{hanisch80}). The horizontal \lq\lq error bars\rq\rq\ in 
this representation are an expression of spread in size of the emission when
measured along the E-W and N-S directions. Superimposed on the data are 
three models computed using the one-dimensional propagators. Clearly, 
the (almost) spherical geometry of the cluster is important for
$r > 1000\,{\rm kpc}$. However, the model should represent 
the data fairly well within the range $500<r<1000$. It is apparent from
this figure that each of the three types of transport can produce the required
fall off with distance, but that the associated predicted 
softening in the spectral index, which is also shown in the figure, is 
quite different in each case. The spectral index maps of 
Giovannini~et~al.~(\cite{giovanninietal93}), which were computed from the 
ratio of the fluxes at 326 and 1380~MHz, show that within a few hundred 
kpc of the edge (at 500~kpc) the spectrum softens rapidly -- by more than 
0.6 in 200~kpc. This is indicated by an upper limit in the figure. 
Of the curves
presented in Fig.~\ref{coma1d}, only that corresponding to supra-diffusion
produces a comparable effect. Thus, in the stationary state, and assuming 
{\it in situ} acceleration is
negligible,
diffusion is not capable of propagating the electrons outwards
from the inner $500\,{\rm kpc}$ of the cluster.

The conclusion that anomalous transport is necessary is based on 
assuming that both the transport coefficients
and the magnetic field (and also 
the type of transport as determined by $\alpha$) are
not functions of position, and further, that the transport is independent of 
particle energy. Although it is not straightforward to 
include such effects in a transport theory, they do not provide
a simple explanation of the observations shown in Fig.~\ref{coma1d} in terms
of standard diffusive transport. 
For example, a magnetic field which decreases with distance from the core,
could explain the fall-off in the intensity, 
but could not account for the observed softening of the 
spectrum, since energy losses would then be less important at 
larger radius than assumed in the figure. Similarly, if the spatial diffusion
coefficient increases in proportion to particle energy -- as expected in the 
case of gyro-Bohm diffusion -- then particles responsible for higher frequency
radiation are more mobile than those radiating at lower frequency. In this
case, the intensity should fall off more slowly at higher frequency, 
which would tend to produce a harder spectrum at larger radius, contrary to 
the observations.
\section{Conclusions}
In this paper we have presented a simple one-parameter method of 
modelling the effects of anomalous transport on energetic electrons.
Anomalous transport is likely to occur wherever magnetised electrons 
(i.e., electrons which are tied to field lines) move in a magnetic field 
which has a stochastic component. In astrophysics, this is the rule 
rather than the exception and anomalous transport can be expected,
for example, in the interstellar medium, as well as 
in the intra-cluster medium of clusters of galaxies. 
The primary observational diagnostic is the intensity and spectrum of 
the synchrotron radiation emitted by the transported particles. 
For the simplest case in which the magnetic field is of constant 
magnitude and of random orientation, we have presented general 
expressions for the surface 
brightness as a function from position of injection, and also for the 
spatial variation of the spectral index.

Applying these to the diffuse emission observed from the outer parts of 
the Coma cluster, we note that it is not possible to distinguish between 
the various forms of transport merely from the profile of the surface 
brightness. However, the expected spatial dependence of the spectral index
is very sensitive to the type of transport. Assuming that the effects of 
particle acceleration are negligible in the outer parts of the cluster, 
and that the electron distribution has achieved a steady state, we find 
that standard diffusive transport cannot produce the observed rapid 
softening of the  spectrum with radius. Under these assumptions, 
the type of transport indicated is supra-diffusion, in which particles 
move almost ballistically in a field configuration which has an ordered 
radial component.

The computations we have presented contain several major 
simplifications. In addition to the assumption of constant, randomly 
orientated magnetic field, and the simple planar or spherical geometry,
we have assumed that the parameters governing the transport are 
independent of the particle's energy. In reality, the type of transport 
itself (i.e., the value of $\alpha$) will change depending on the energy 
range and timescales considered. Thus, at very large times (which may 
exceed the synchrotron lifetime), a particle can be expected to 
decorrelate from the magnetic field and perform diffusion (e.g., 
Duffy et al.,~\cite{duffyetal95}). We do not model the situation in which a 
significant change in $\alpha$ occurs within the synchrotron lifetime of 
an electron. Finally, in order to apply such models to well-observed 
objects such as spiral galaxies, it will be 
necessary to include additional effects such as 
that of a galactic wind, as well as bremsstrahlung and ionisation loss 
processes. However, these processes will not change our basic conclusion 
that it is the spatial dependence of the synchrotron spectral index 
which provides the most sensitive measure of the transport properties of the 
emitting electrons.

\noindent{\bf Acknowledgements:}
BRR thanks the Max-Planck-Institut f\"ur Kernphysik for the grant of 
a visitor's stipend, during which this work was performed. We are grateful
to R.O.~Dendy, P.~Duffy and Y.A.~Gallant for stimulating discussions.

\appendix
\section{Derivation of the propagators}
The propagators are evaluated by inserting the expansions (\ref{EXPS})
into Eq.~(\ref{MWE}), inverting the Fourier transformation analytically 
(since we fix $\beta=2$) and using the method of steepest descents to 
find the asymptotic dependence of the $\Xi(\xi)$ at large $\xi$.
For the one-dimensional case, a standard integral leads to
\eqb
P(x,s)&=&{\xi\over s}\exp\left[-\xi(s\,t)^{\alpha/2}\right]
\eqe
with $\xi=\sqrt{2}(x/\sigma)(\tau_D/t)^{\alpha/2}$,
from which, by the method of steepest descents (e.g., Mathews \& 
Walker~\cite{mathewswalker70}) one finds
\eqb
P(x,t)&\approx&{1\over\sigma\sqrt{2\pi(2-\alpha)}}\left({\tau_D\over 
t}\right)^{\alpha/2}
\left({\alpha\xi\over2}\right)^{-(1-\alpha)/(2-\alpha)}
\nonumber\\
&&\exp\left[ -\left(2-
\alpha\over\alpha\right)\left({\alpha\xi\over2}\right)^{2/(2-
\alpha)}\right]
\eqe
In the case of diffusion ($\alpha=1$), this result is correctly
normalised, i.e., $\int_{-\infty}^{\infty}
\diff x\,P(x,t)=1$. However, in general the normalisation is lost
in this approach (cf.~Rax \& 
White~\cite{raxwhite92}). Correcting for this, we find for the constants 
used in Eq.~(\ref{xioned}) the expressions
\eqb
\label{constants1}
A&=&\left({2-\alpha\over\alpha}\right)\left({\alpha\over2}\right)^{2/(2-
\alpha) }\\
\label{constants2}
C_d&=&
{1\over2-\alpha}\left[{d(2-\alpha)\over\pi}\left({\alpha\over2}\right)^{\alpha/(2-\alpha)}
\right]^{d/2}
\eqe

\section{The synchrotron brightness distribution}

The expression for the density (\ref{statdens}) can be 
directly integrated numerically. Alternatively, it is easily computed
by expanding the factor 
$(1-g \gamma t')^{p-2}$ and integrating over $t'$, which yields 
\eqb
n(\vec{x},\gamma)&=&{Q_0\gamma^{-p-1} \over
			\alpha \pi^{d/2}g |\vec{x}|^d}
			\sum_{k=0}^{+\infty} h_k(p-2) 
		\left({ |\vec{x}| \over R(\gamma)}\right)^{{2 \over \alpha} (k+1)} 
		\nonumber \\ 
		& & \times  
		\Gamma\left({d \over 2} - (k+1) {2-\alpha \over \alpha}, 
		\; \left({ |\vec{x}| \over R(\gamma)}\right)^{{2 \over 2-\alpha}} 
		\right) 
\label{seriestatdens}
\eqe
where $$R(\gamma) = {2 x_s \over \alpha \sqrt{d}} 
\left({\alpha \over 2-\alpha }\right)^{{2-\alpha \over 2}} $$ 
and $$ h_0(n)=1 \; , \; \; h_k(n)={(-1)^k \over k!} n(n-1)...(p-k+1) \; .$$
Note that for $p$ integer and larger than 2, the series reduces to 
$p-2$ terms. \\ 

This expression recovers, for 3-dimensional 
diffusion ($\alpha=1$, $d=3$), the formula derived by Wilson 
(\cite{wilson75}), and used by Valtaoja (\cite{valtaoja84}) 
\eqb
n(r,\gamma)&=& { Q_0 \sqrt{g} \Gamma(p-1) \gamma^{-p+1/2} \over
8 \pi^{3/2} D^{3/2} } 
\exp\left(- {g \gamma r^2 \over 4 D} \right) 
\nonumber\\
&&
U\left(p-1;{3 \over 2};{g \gamma r^2 \over 4 D} 
\right) \; , 
\label{Wilsonstatdens}
\eqe  
where $U$ is the confluent hypergeometric function of the second kind 
(Abramowitz and Stegun~\cite{abramowitzstegun72}). 
This can be seen either directly from Eq.~(\ref{statdens}) or,
in the special case $p=2$, by noting that
\eqb 
\sqrt{X} e^{-X} U\left(1;{3 \over 2};X \right) &=& \Gamma\left(
{1 \over 2},X \right) \; .
\eqe

\end{document}